\documentclass[aip,jcp,reprint]{revtex4-1}
\usepackage{graphicx}
\usepackage{color}
\usepackage{amsmath}
\usepackage{amssymb}
\usepackage{anysize}

\begin{document}

\title{Coarse-grained modelling of supercoiled RNA}
\author{Christian Matek}
\affiliation{Rudolf Peierls Centre for Theoretical Physics, University of Oxford, 1 Keble Road, Oxford, OX1 3NP, United Kingdom}
\author{Petr \v{S}ulc}
\affiliation{Center for Studies in Physics and Biology, The Rockefeller University, 1230 York Avenue, New York, NY 10065, USA}
\author{Ferdinando Randisi}
\affiliation{Rudolf Peierls Centre for Theoretical Physics, University of Oxford, 1 Keble Road, Oxford, OX1 3NP, United Kingdom}
\affiliation{Life Sciences Interface Doctoral Training Center, South Parks Road, Oxford, OX1 3QU, United Kingdom}
\author{Jonathan P. K. Doye}
\affiliation{Physical and Theoretical Chemistry Laboratory, University of Oxford, South Parks Road, Oxford, OX1 3QZ, United Kingdom}
\author{Ard A. Louis}
\affiliation{Rudolf Peierls Centre for Theoretical Physics, University of Oxford, 1 Keble Road, Oxford, OX1 3NP, United Kingdom}

\begin{abstract}
We study the behaviour of double-stranded RNA under twist and tension using oxRNA, a recently developed coarse-grained model of RNA.
Introducing explicit salt-dependence into the model allows us to directly compare our results to data from recent single-molecule experiments.
The model reproduces extension curves as a function of twist and stretching force, including the buckling transition and the behaviour of plectoneme structures.
For negative supercoiling, we predict denaturation bubble formation in plectoneme end-loops, suggesting preferential plectoneme localisation in weak base sequences.
OxRNA exhibits a positive twist-stretch coupling constant, in agreement with recent experimental observations.
\end{abstract}

\maketitle

\section{Introduction}
Due to their importance in the storage and processing of genetic information, 
nucleic acids play a fundamental role in many biological processes such as transcription, translation and replication.\cite{Alberts2007,Elliott2011}
In their double stranded (ds) form, DNA and RNA adopt a helical geometry. While dsDNA typically forms a B-helix, dsRNA adopts an A-helical form, which is wider, has a smaller pitch and bases that are inclined with respect to the helical axis.\cite{neidle2010principles}
Double-stranded DNA and RNA exhibit complex mechanical behaviour that is important in many biomechanical contexts, such as genome organisation,\cite{Kouzine2013} virus packaging~\cite{Patton2000,Guo2007} and nucleosome positioning.\cite{Andrews2011}

Moreover, both DNA~\cite{Seeman2010,Zhang2014} and more recently RNA~\cite{Guo2010} have emerged as versatile building materials on the nanoscale.
Driven by these wide-ranging applications, the mechanical properties of nucleic acids have been studied with increasing precision on a single-molecule level.\cite{Kapanidis2009}
While the mechanical behaviour of dsDNA has been widely characterised using molecular tweezer assays,\cite{Strick1996,Bustamante2003,Brutzer2010,Mosconi2009,Forth2008,Janssen2012,Loenhout2012,Vlaminck2012} dsRNA has received less attention.\cite{Abels2005,herrero2012mechanical}
The first comprehensive experimental study of the twisting and stretching behaviour of dsRNA was only recently carried out by Lipfert and co-workers.\cite{Lipfert2014}

Correspondingly, theoretical work using atomistic simulations,\cite{Orozco2008,Liverpool2008} continuum models~\cite{Neukirch2011,Daniels2011} and coarse-grained simulations~\cite{Ouldridge2011,Matek2015,Chou2014} has centered on modelling the properties of torsionally stressed DNA.
There have been far fewer studies of supercoiled dsRNA, although theoretical investigations exist using atomistic simulations\cite{Wereszczynski2006} and the HelixMC package, which uses a base-pair-level description of the molecule.\cite{Chou2014}

Here, we study the behaviour of supercoiled dsRNA using a salt-dependent extension of oxRNA, a recently developed nucleotide-level model of RNA.\cite{Sulc2014,Sulc2014a} The model is developed to capture the structural, mechanical and thermodynamical properties of both single-stranded and double-stranded RNA and was previously used to study RNA hairpin unzipping, the thermodynamics of pseudoknot folding, kissing complex formation and toehold-mediated strand displacement.\cite{Sulc2014,vsulc2014modelling}
The coarse-graining methodology of oxRNA allows us to capture the effects of double-strand denaturation, which are not accessible in continuum or basepair-level models. 
Likewise, the computational efficiency gained by the coarse-graining allows us to access time scales and system sizes relevant to the physics of double-strand buckling and denaturation, which are currently beyond the scope of all-atom molecular dynamics simulations.  
We previously used a coarse-grained model of DNA, oxDNA \cite{Ouldridge2011,oxDNA}, to study the supercoiling of dsDNA and obtained good agreement with experimental results.\cite{Matek2015,Matek2012} 
In this work, we use oxRNA to directly compare to a recent experimental study of dsRNA supercoiling.\cite{Lipfert2014}

This paper is organised as follows. First, we briefly describe an extension of the oxRNA model to include a salt-dependent parameterisation. 
We then compare the model prediction to recent measurements of the end-to-end distance and torque response of dsRNA as a function of imposed stretching force and superhelical density.\cite{Lipfert2014}
We extract parameters characterising the twisting, bending and extensional behaviour of the molecule.
The results of our simulations are in reasonable agreement with experimental data. 
For negative supercoiling and intermediate stretching forces, we observe denaturation bubble formation localised in plectoneme end-loops, similarly to what was found in a previous work on DNA plectonemes using a related modelling approach for DNA.\cite{Matek2015}

\section{OxRNA model with salt-dependent interaction}

\begin{figure}[tb]
\includegraphics[width=0.55\textwidth]{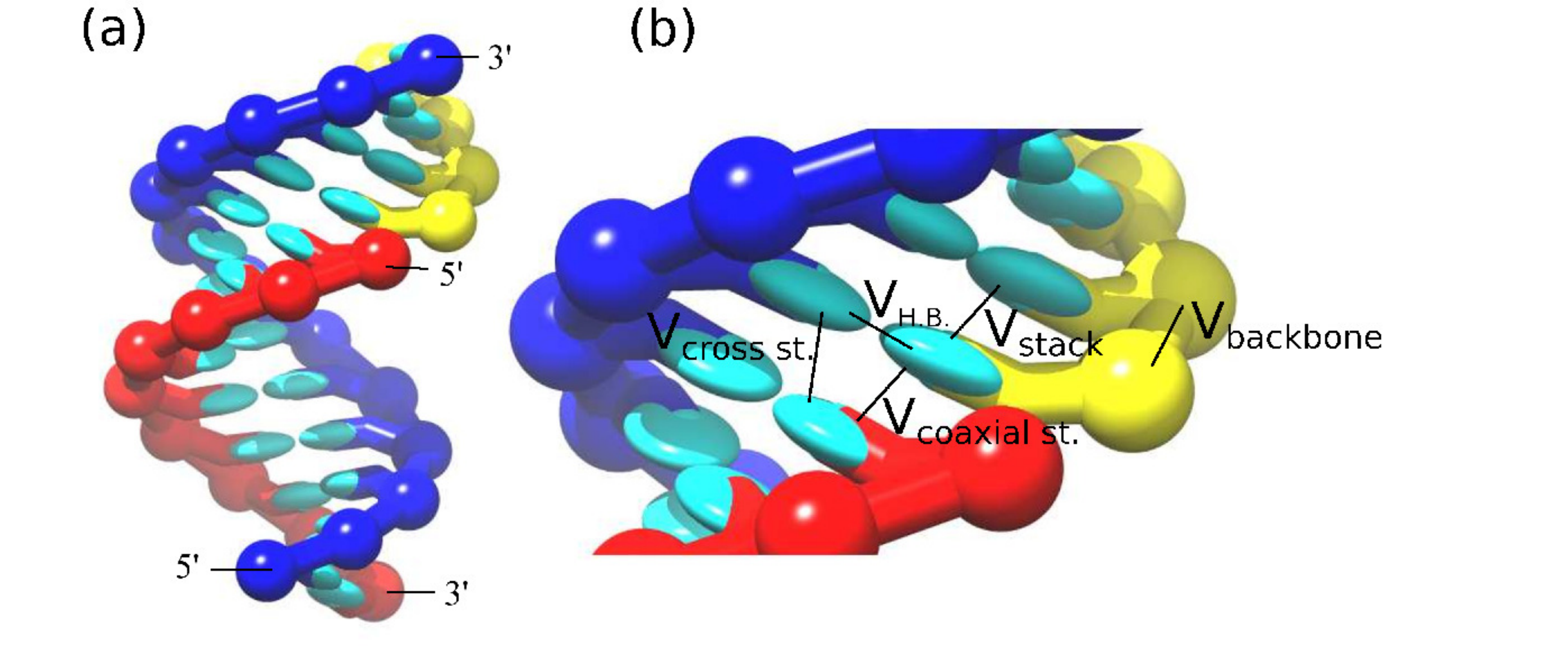}
\centering
\caption{A schematic representation of (a) an A-RNA helix as represented by the oxRNA model and (b) the attractive interactions in oxRNA. The lines in (b) schematically show the interactions between the nucleotides: Hydrogen bonding ($V_{\rm{H.B.}}$), stacking ($V_{\rm{stack}}$), cross-stacking ($V_{\rm{cross~st.}}$) between a nucleotide and the nucleotide that is the $3^{\prime}$ neighbour of the directly opposite nucleotide and coaxial stacking ($V_{\rm{coaxial~st.}}$). The nucleotides also interact with excluded-volume interactions and electrostatic interactions, which are not shown.}
\label{fig_potentials}
\end{figure}

OxRNA represents each nucleotide as a single rigid body with multiple interaction sites. The rigid bodies interact with effective anisotropic interactions that are designed to
capture the overall thermodynamic and structural consequences of the base-pairing, stacking and backbone interactions, as schematically shown in Fig.~\ref{fig_potentials}.
The potential of the oxRNA model is 
\begin{eqnarray}
\label{eq_potential}
  V_{\rm oxRNA} &=&  \sum\limits_{\left\langle ij \right\rangle}  \left( V_{\rm{backbone}} + V_{\rm{stack}} + V^{'}_{\rm{exc}} \right)   \nonumber \\
                &+& \sum\limits_{i,j \notin {\left\langle ij \right\rangle}} \left( V_{\rm{H.B.}}  +  V_{\rm{cross~st.}}  + V_{\rm{exc}}  \right.  \nonumber \\ 
                &+& \left. V_{\rm{coaxial~st.}} + V_{\rm{electrostatic} } \right), 
\end{eqnarray}
where the first sum runs over all pairs of nucleotides which are nearest neighbours on the same strand and the second sum runs over all other pairs. A detailed description of the interactions and their parameterisation is provided in Ref.~\onlinecite{Sulc2014}, with the exception of $V_{\rm{electrostatic}}$ which is newly introduced to explicitly capture salt-dependent effects. This term is isotropic and is centred on the backbone site of each nucleotide. The functional form of the potential is based on Debye-H\"uckel theory, where we further introduce a cutoff at a finite distance.
We use the Debye-H\"uckel length for water and treat the strength of the effective negative charge on the backbone site as a parameter, which we fit 
to reproduce the melting temperatures of duplexes of lengths 5, 6, 7, 8, 10 and 12 at salt concentrations varying from $0.1\,{\rm M}$ to 1\,M.
To obtain the melting temperatures to which we fit,
we use the averaged nearest-neighbour model of Turner {\it et al.} \cite{mathews1999expanded} extended with a salt-dependent free-energy correction inferred from hairpin unzipping experiments
at varying salt conditions.\cite{stephenson2013combining} We employ the fitting procedure based on thermodynamic integration, as detailed in Ref.~\onlinecite{Snodin2014}.
We provide further details of the functional form of $V_{\rm{electrostatic}}$ and its parameterisation in the Supplementary Material.\footnote{see Supplementary Material for further data and details of the simulation setup and model parameterisation.}

The backbone interaction, $V_{\rm{backbone}}$, is an isotropic FENE spring potential
that is used to mimic the covalent bonds in the RNA backbone that constrain the intramolecular distance between neighbouring nucleotides. The nucleotides further
have repulsive excluded-volume interactions $V_{\rm{exc}}$ and $V^{'}_{\rm{exc}}$ that depend on the distance between their interaction
sites, namely the backbone-backbone, stacking-stacking and stacking-backbone distances.
The excluded-volume interactions ensure that strands cannot overlap, or pass through each
other in a dynamical simulation.

The duplex is stabilised by hydrogen bonding ($V_{\rm{H.B.}}$), stacking
($V_{\rm{stack}}$) and  cross-stacking ($V_{\rm{cross~st.}}$) interactions.
These potentials are anisotropic and depend on the distance between the relevant
interaction sites as well as the mutual orientations of the nucleotides.
The hydrogen-bonding term $V_{\rm{H.B.}}$ captures the stabilising interactions between complementary Watson-Crick (AU and GC) and wobble (GU) base pairs, while $V_{\rm{stack}}$ mimics the favourable interaction between adjacent bases on the same strand. The strength of $V_{\rm{H.B.}}$ and $V_{\rm{stack}}$ is sequence-dependent, i.e. depends on the identity of the interacting bases.  

The cross-stacking potential, $V_{\rm{cross~st.}}$, is designed to capture the
interactions between diagonally opposite bases in a duplex and has its minimum when the distance and mutual orientation
between nucleotides corresponds to that for a nucleotide and the
3$^\prime$ neighbour of the directly opposite nucleotide in an A-form helix. This interaction has been parameterised to capture the stabilisation of an RNA duplex by a 3$^{\prime}$ overhang.

The coaxial stacking potential $V_{\rm{coaxial~st.}}$  represents the
stacking interaction between nucleotides that are not nearest neighbours on the
same strand.

In this work, we use the average-base parameterisation of oxRNA, which only allows for specific formation of AU and GC Watson-Crick base pairs.
Hydrogen-bonding energies between complementary base-pairs and stacking energies are set to identical, average strengths.
This choice allows us to focus on the generic properties of RNA double strands, which are independent of specific sequence properties.
Parameters are fitted to reproduce the thermodynamics of hairpins and duplexes averaged over all possible combinations of Watson-Crick base pair steps, as predicted by the model of Turner and collaborators.\cite{mathews1999expanded}
We note that the model cannot reproduce tertiary structure contacts such as ribose zippers or Hoogsteen base pairs, 
but we do not anticipate that these non-canonical interactions will be relevant for the modelling of the behaviour observed in Ref.~\onlinecite{Lipfert2014}.

\section{Simulation methods}
\label{sec:methods}
The results reported in this work were obtained from molecular dynamics simulations of oxRNA using an Andersen-like thermostat (described in the appendix of Ref.~\onlinecite{Russo09}) at 300\,K using both the CPU and GPU implementation of the model.\cite{rovigatti2014comparison}
We intentionally set the diffusion constant artificially high to speed-up convergence of the simulations to equilibrium. In particular, we used the translational diffusion constant $D = 5.8 \times 10^{-7}$\,m$^2$s$^{-1}$, which
corresponds to a diffusion constant of $2.1 \times 10^{-8}$\,m$^2$s$^{-1}$ for a 14-mer and is about two orders of magnitude more than the experimentally measured $D_{\rm exp} = 0.92 \times 10^{-10}$\,m$^{2}$s$^{-1}$.\cite{Lapham1997}
The simulation time step was set to $1.22 \times 10^{-14}\,{\rm s}$. 

\begin{figure}[tb]
\includegraphics[width=0.3\textwidth]{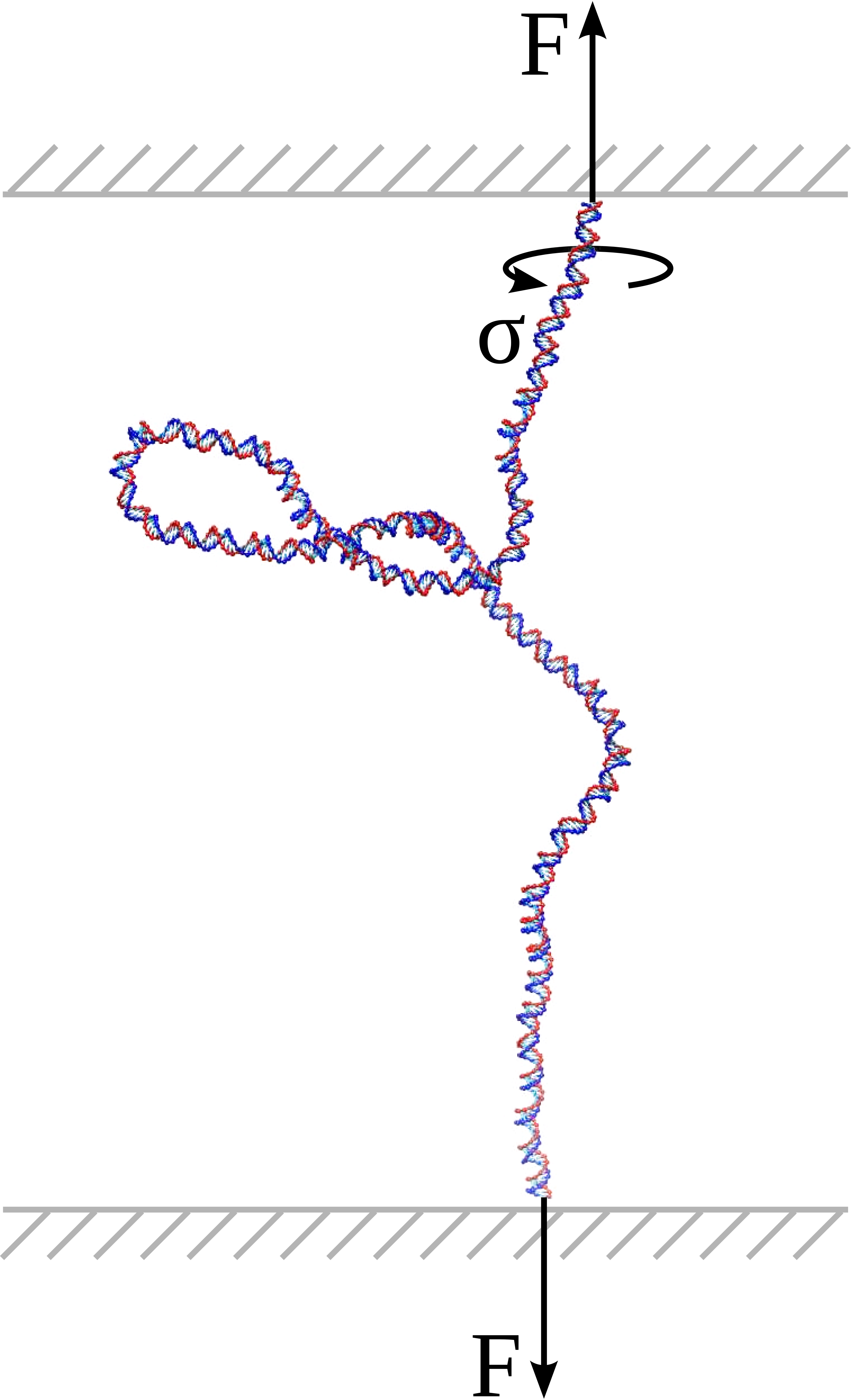}
\centering
\caption{Schematic of the simulation setup used in this work. A-helical dsRNA strands are subjected to torsional stress by fixing a constant superhelical density $\sigma$ and exerting a stretching force $F$ to the strand ends. Repulsive planes tagged to the strand ends (indicated in grey) ensure the superhelical density remains constant by preventing the duplex from passing around its own ends. The configuration shown was obtained in a simulation at $\sigma = +0.08$ and $F=3.0$~pN. Under these conditions, a plectoneme forms leading to significant shortening of the end-to-end extension.}
\label{Fig:setup}
\end{figure}

We simulated 600-bp dsRNA molecules using an average-base parameterisation of oxRNA that includes base-pair specificity, but ignores sequence-dependent variations in interaction energies.\cite{Sulc2014}
The duplex was set-up as a homogenously twisted helix with a desired superhelical density and pre-equilibrated for a simulation time of at least 1\,$\mu$s. Simulations were then run for at least 8\,$\mu$s of simulation time. 
The superhelical density is defined as $\sigma = p_0/p - 1$, where $p$ is the imposed pitch and $p_0$ is the equilibrium pitch of dsRNA when no stress is applied. 
To keep superhelical densities constant during a simulation run, strand ends were fixed in two-dimensional harmonic traps and the strands prevented from passing around their own ends, as described in detail in the Supplementary Material.\cite{Note1}
The resulting setup of the dsRNA systems subject to linear and torsional stress is illustrated schematically in Fig.~\ref{Fig:setup}. 

To match the experimental conditions of Ref.~\onlinecite{Lipfert2014}, all simulations were run at a monovalent salt concentration of 100\,mM.

\section{Results}

Superhelical stress can be stored in dsDNA and dsRNA by both twisting and writhing.
For small values of supercoiling, the torsional energy of the system grows until a buckling superhelical density $\sigma_b$ is reached, at which it becomes more favourable for the system to form writhed structures known as plectonemes (see Fig.~\ref{Fig:setup}) where the supercoiling energy is stored in bending rather than twisting.\cite{Strick2003}
Writhing, which results in a shortening of the molecule end-to-end distance, is disfavoured by applying an external stretching force.
As described in more detail below, our model exhibits this generic behaviour, as expected for a twist-storing polymer with finite bending persistence length.
Here we compare the behaviour of our model to experimental data of Lipfert and co-workers.\cite{Lipfert2014}

\begin{figure*}
 \includegraphics[width=\textwidth]{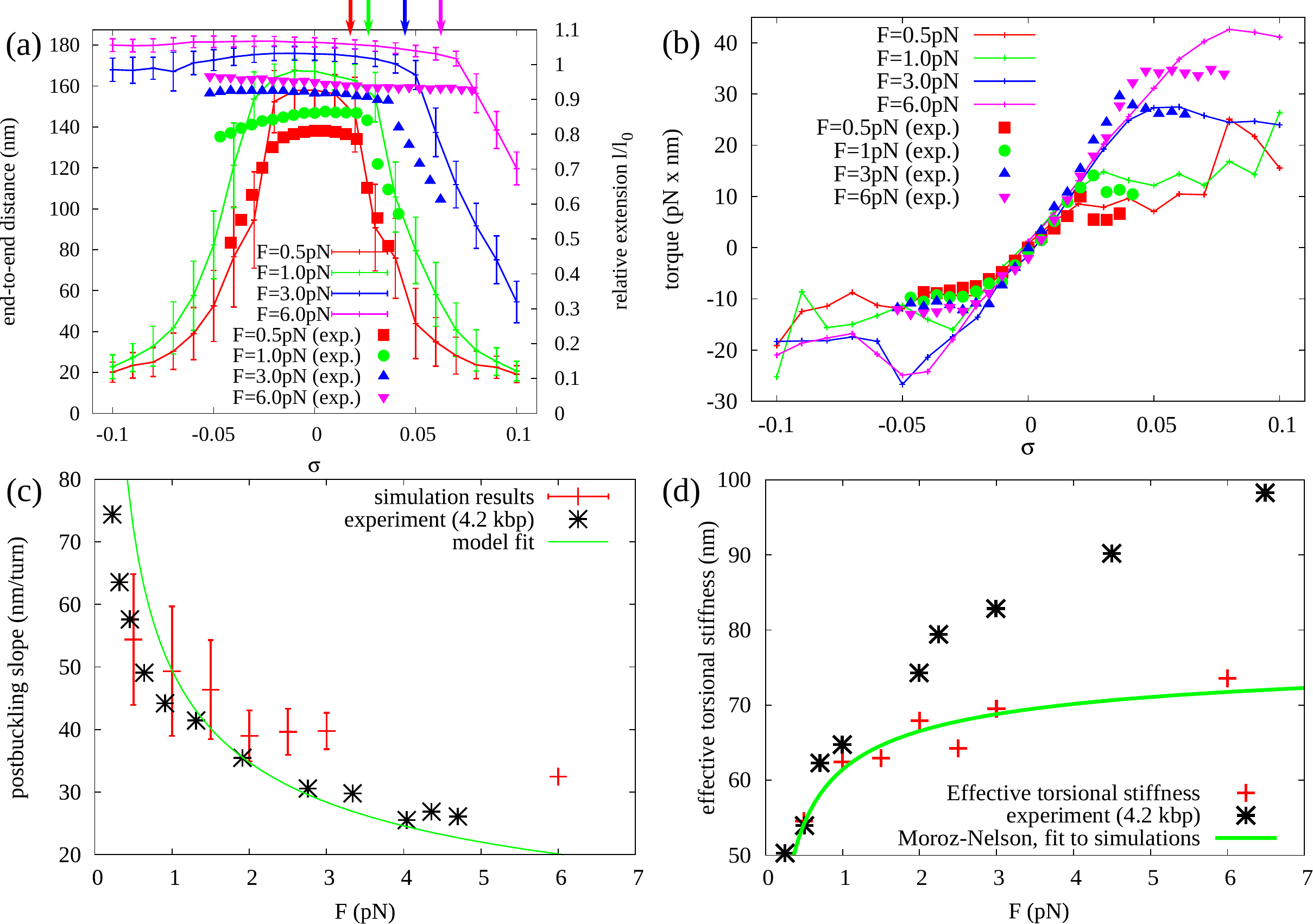}
 \caption{Mechanical behaviour of 600-bp dsRNA under torsional stress in the oxRNA model at 100mM monovalent salt, together with experimental data obtained for a 4.2\,kbp-system from Ref.~\onlinecite{Lipfert2014}: (a) ``Hat curves'' showing the end-to-end extension of the system. Arrows indicate the values of $\sigma$ for which buckling is expected from Eq.~\ref{eq:buckling_sigma}. Error bars (standard deviations) indicate the magnitude of thermal fluctuations rather than measurement uncertainties. (b) Torque response of the dsRNA strand, showing a linear regime at low twist, followed by a constant-torque regime after buckling. (c) Postbuckling slopes measured from the hat curves in (a) and a fit of simulation results to the analytical model of Ref.~\onlinecite{Marko2007}, leading to a torsional stiffness of the plectonemic state $P_{\rm oxRNA}=22$\,nm. (d) Fit of the torsional stiffness measured in simulations to a Moroz-Nelson model.\cite{Moroz1997} }
 \label{Fig:Mechanics}
\end{figure*}

\subsection{Force-extension response at varying superhelical densities}

\begin{figure}[tb]
\includegraphics[width=0.4\textwidth]{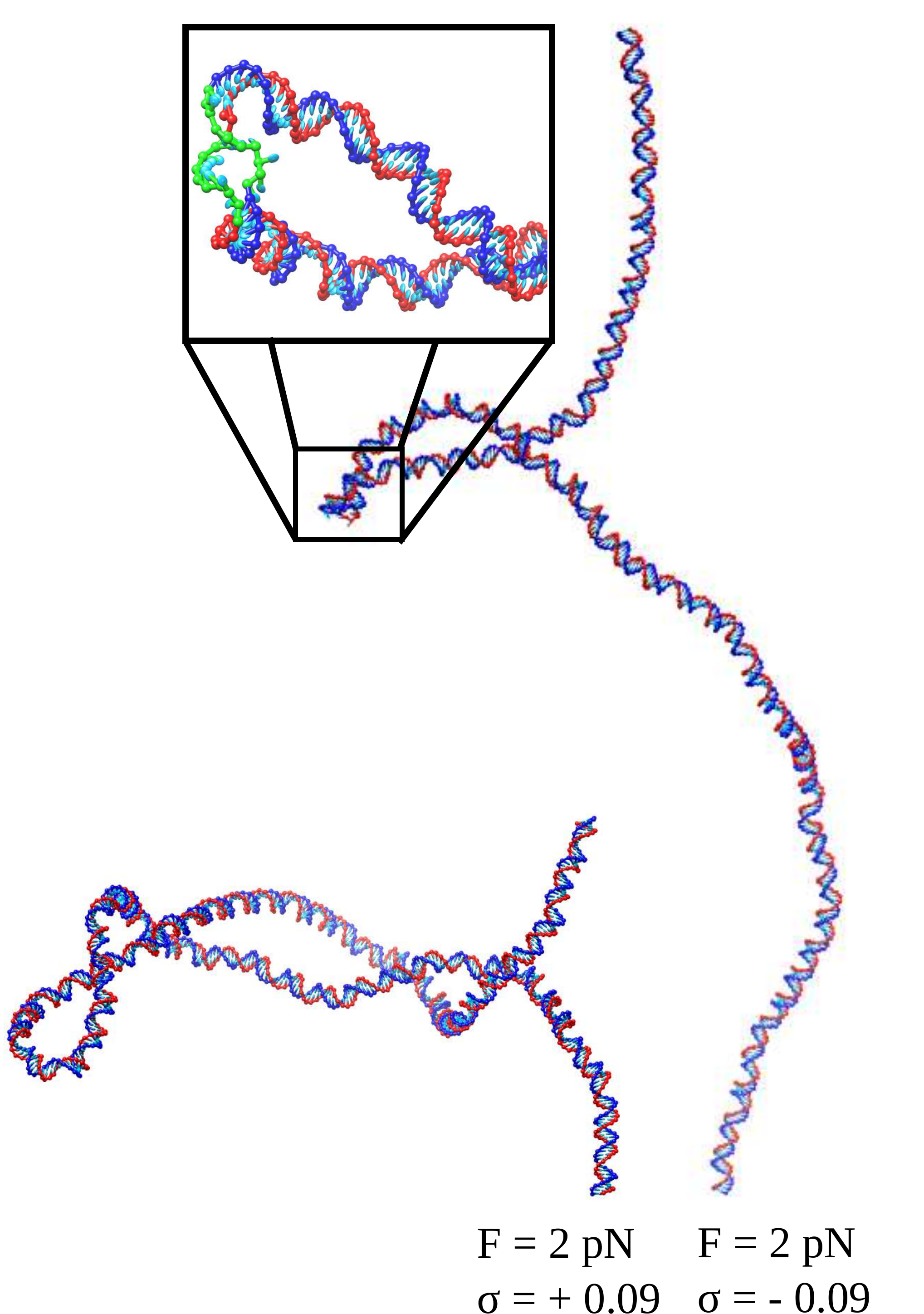}
\centering
\caption{Strand configurations observed at a stretching force $F=2.0$\,pN for $\sigma = +0.09$ (left) and $\sigma = -0.09$ (right). At negative superhelical density, a denaturation bubble of size 10\,bp is observed in the end-loop of the plectoneme, analogous to the behaviour predicted for dsDNA in Ref.~\onlinecite{Matek2015}. Enlarged structures show the microscopic configuration of the plectoneme end-loop, where denatured bases are coloured green.}
\label{Fig:tip_bubble_plect}
\end{figure}

We first study the end-to-end extension of a 600-bp dsRNA as a function of superhelical density $\sigma$ and stretching force $F$. 
For a given dsRNA with imposed $\sigma$ and $F$, we run a molecular dynamics simulation, as described in Section \ref{sec:methods}, and measure the end-to-end distance between the first and the last base pairs of the duplex. 
The results are shown in Fig.~\ref{Fig:Mechanics}(a).
When the superhelical density of the dsRNA molecule in our model is increased, its end-to-end extension initially changes little, until a buckling point is reached at which it is thermodynamically more favourable for the system to bend into a plectonemic structure than to further twist. 
For stretching forces $F\gtrsim 2$\,pN, the extension curves become asymmetric, as denaturation rather than plectoneme formation occurs for negative supercoiling (Fig.~\ref{Fig:tip_bubble_plect}). 

Comparison of our simulation results to the recent experimental data of Ref.~\onlinecite{Lipfert2014} (included in Fig.~\ref{Fig:Mechanics}(a)) shows good agreement for the buckling superhelical densities, post-buckling slopes, and the onset of double-strand melting, indicating that the overall behaviour of dsRNA subject to twist and stretching force is well reproduced by oxRNA. 

Nevertheless, oxRNA still buckles under positive supercoiling for stretching forces above 5\,pN, while no buckling was observed in experiment above such a force.\cite{Lipfert2014} It was proposed that overwound dsRNA above 5\,pN changes its conformation to a  ``P-RNA'' state that is similar to the P-DNA structure of dsDNA, which is characterised by interwound sugar-phosphate backbones with exposed bases.\cite{allemand1998stretched} Such a structure is not observed with oxRNA under these conditions.  

Compared to the experimental data, simulated dsRNA molecules show a larger relative end-to-end extension. This is due to the relatively low value of the extension modulus $K_{\rm oxRNA}\approx116$\,pN in oxRNA, which is significantly lower than the experimental value of $K_{\exp}\approx350$\,pN.\cite{Lipfert2014}
However, for sufficiently low forces, the buckling behaviour of the strand is expected to be only minorly affected by this discrepancy.

As was done in the experimental study, we further determined the twist-stretch coupling by measuring the slope of the end-to-end extension curve at low superhelical densities ($-0.02 \leq \sigma \leq +0.025$) and high stretching force $F=6.0$\,pN. 
We obtain a twist-stretch coupling of $(d\Delta L/d Lk)_{\rm oxRNA}=-0.72$\,nm/turn, which is to be compared to an experimental value of $(d\Delta L/d Lk)_{\rm exp}=-0.85$\,nm/turn.\cite{Lipfert2014} 
Thus, oxRNA qualitatively reproduces the positive twist-stretch coupling observed for RNA.

To further quantify the mechanical behaviour of oxRNA, we measured the slopes of the extension curves in the postbuckling regime (shown in Fig.~\ref{Fig:Mechanics}(c)). 
We note that the values obtained are sensitive to the selection of points included in the fit of the postbuckling slope, as indicated by the error bars in Fig.~\ref{Fig:Mechanics}(c).
The fitting procedure is described in detail in the Supplementary Material.\cite{Note1}
Again, approximate agreement with experimental values is found. 
When fitting to a thermodynamic model of the plectonemic phase,\cite{Marko2007} qualitatively similar but more pronounced systematic deviations occur compared to the experimental data, as shown in Fig.~\ref{Fig:Mechanics}(c).
At least part of the discrepancies may be due to finite size effects in the simulated 600-bp system, which approximates the thermodynamic limit less well than the 4.2-kbp experimental system does.\cite{Matek2015,Note1}

\begin{figure*}
 \includegraphics[width=.99\textwidth]{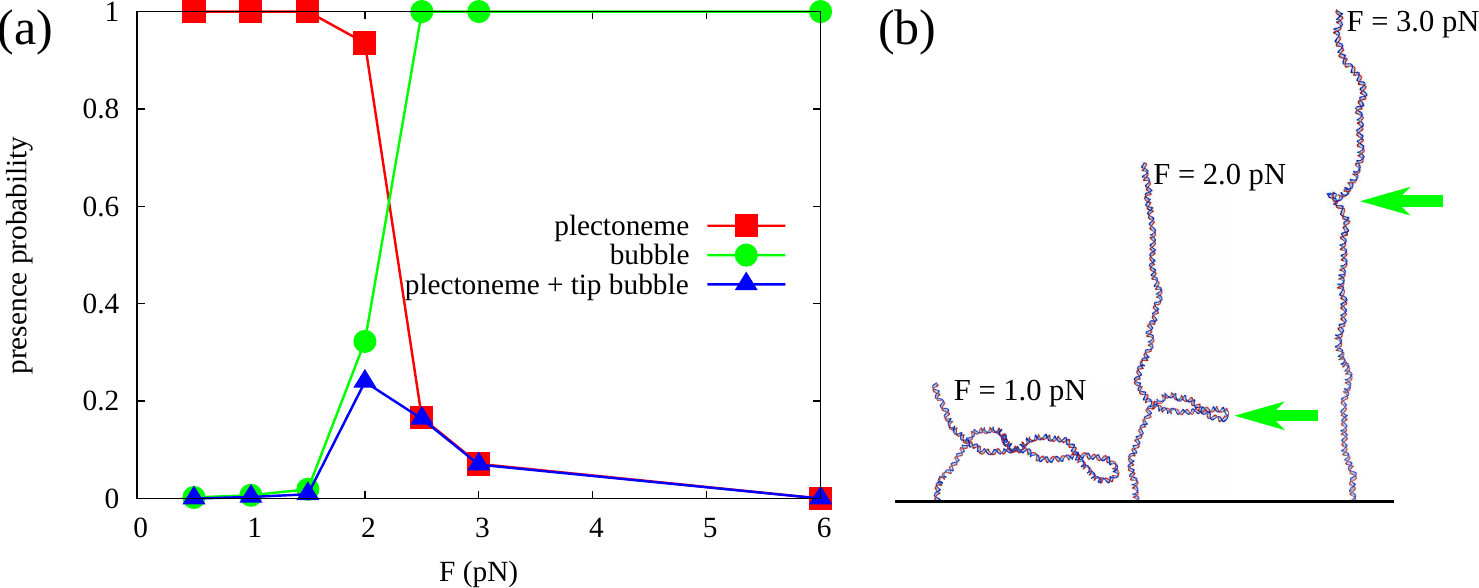}
 \caption{Characterising the tip-bubble state in dsRNA plectonemes: (a) Frequency of plectoneme, co-localised tip-bubble and pure bubble states as a function of applied stretching force at $\sigma=-0.09$. (b) Representative RNA configurations at $\sigma=-0.09$ and different stretching forces, with bubble positions indicated by green arrows. At $F=1.0$\,pN, no stable strand denaturation occurs, while a 5-bp denaturation bubble localised at the plectoneme tip is found at $F=2.0$\,pN, and a pure, writhed denaturation bubble of size 20\,bp occurs at $F=3.0$\,pN.}
 \label{Fig:Writhing_Melting}
\end{figure*}

As in a recent study on dsDNA using the oxDNA model,\cite{Matek2015} 
we observed localisation of double-strand denaturations in the end-loop of plectoneme structures 
(see Figs.~\ref{Fig:tip_bubble_plect} and \ref{Fig:Writhing_Melting} for $\sigma < 0$ and intermediate stretching force $F\approx 2$\,pN~\cite{Note1}).
In this configuration, the enthalpic cost for opening the bubble is partially compensated by the lower bending energy of a plectoneme end-loop containing a denaturation bubble; the bubble also reduces the torsional stress by absorbing negative twist.

We note however that the prevalence of these bubbles co-localised in the end loops of the plectonemes is reduced compared to the analogous setup in dsDNA.
Primarily, this difference may be attributed to the stronger base-pairing of Watson-Crick base pairs in RNA compared to DNA,\cite{xia1998thermodynamic} making bubble opening in stressed parts of the strand more enthalpically costly.
More subtle effects, such as differences between the A-form helical geometry of dsRNA and the B-form helical geometry of dsDNA, as well as details of the model of screened electrostatic interactions may further contribute to the differences observed.
We also note that the simulations presented in this work were done at $0.1$\,M monovalent salt rather than the $0.5$\,M used in Ref.~\onlinecite{Matek2015} with oxDNA.
At variance with the dsDNA case,\cite{Matek2015} we observed double strand denaturation only for $\sigma<0$ (Fig.~\ref{Fig:tip_bubble_plect}), while no significant denaturation occurred for positive supercoiling at the stretching forces studied in this work. This may again be explained by the stronger base-pairing free-energy in dsRNA. 
Force-induced melting of the duplex is expected also for $\sigma>0$ at forces significantly higher than the ones used in this work or in experimental assays.\cite{Herrero-Galan2013}

In this work, we used an average-base parameterisation of oxRNA. However, stable occurrence of a tip-bubble plectoneme state for $\sigma<0$ and intermediate $F \approx 2$\,pN suggests that in a sequence-dependent scenario, the centres of the plectonemes will be primarily localised to AU-rich regions of the strand at these conditions, because their weaker base paring reduces the cost of bubble formation; this mechanism is described in detail for dsDNA in Ref.~\onlinecite{Matek2015}.
We note that the occurrence of co-localised denaturation and writhing is the consequence of the elastic properties of a chiral, semi-flexible polymer combined with the possibility for the double strand to denature, and is therefore expected to be a robust phenomenon that is largely independent of detailed microscopic properties of the molecule.

\subsection{Torque response and mechanical parameters of dsRNA}
We further quantify the properties of dsRNA by studying the torque response of molecule at different superhelical densities and forces.
The torque response of the simulated system to imposed superhelical density is shown in Fig.~\ref{Fig:Mechanics}(b), and compared to the corresponding experimental data.
Overall, we observe fair agreement with the corresponding experimental values.

For small absolute values of the superhelical density, the torque response of the system grows linearly with $\sigma$.
In this regime, the effective torsional rigidity of the system corresponds to the slope of the torque response curve.
The bending and twist persistence lengths $A_0$ and $C_0$ can be determined by fitting the effective torsional rigidities $C_{\rm eff}$ to a model due to Moroz and Nelson~\cite{Moroz1997} (see Fig.~\ref{Fig:Mechanics}(d)):
\begin{equation}\label{eq:Moroz-Nelson-first}
 C_{\rm eff} = C_0 \left[1-\frac{C_0}{4A_0} \sqrt{\frac{k_B T}{A_0F}} + \mathcal{O}\left( F^{-3/2} \right) \right].
\end{equation}
The fits yield $A_{0,\rm oxRNA}=32$\,nm and $C_{0,\rm oxRNA}=79$\,nm for the simulated system. 
Both values are of the correct order of magnitude, but lie below the values $A_{0,\rm exp}=57$\,nm and $C_{0,\rm exp}=100$\,nm determined from the experimental systems in Ref.~\onlinecite{Lipfert2014}. 
The difficulty of correctly reproducing the persistence length in a coarse-grained model of RNA has been noted before,~\cite{Sulc2014} and has also affected other coarse-grained modelling approaches.\cite{Chou2014}
However, as the relative deviations in the elastic persistence lengths are of similar magnitude, we expect properties that only depend on the ratio of twisting and bending energies, such as twist-induced double-strand buckling to be reproduced more accurately by our model than properties that depend on their values separately.

As $|\sigma|$ is increased, a buckling point is reached at which the system forms a plectoneme structure, thus absorbing supercoiling by writhing rather than further twisting, as discussed previously.
Buckling occurs once the superhelical density exceeds a critical value $\sigma_b$,
which is set by the ratio of $C_0$ and $A_0$.
The critical superhelical density can be estimated by\cite{Strick2003,Matek2014a}
\begin{equation}
 \sigma_b = \sqrt{\frac{2FA_{\rm 0}}{k_{\rm B}T}} \frac{r_0 p_0}{2 \pi C_0},
 \label{eq:buckling_sigma}
\end{equation}
where $F$ is the applied stretching force, and $r_0=0.28$\,nm and $p_0=11.14$\,bp are the equilibrium rise and pitch of the dsRNA helix, respectively.\cite{Sulc2014}
Using the persistence length values obtained by fitting to Eq.~\ref{eq:Moroz-Nelson-first}, $\sigma_b$ can be predicted from Eq.~\ref{eq:buckling_sigma}.
As indicated by arrows in Fig.~\ref{Fig:Mechanics}(a), the critical superhelical densities obtained in this way are consistent with the buckling behaviour observed in simulations.
They are also consistent with experiment, although it should be kept in mind that part of the accuracy arises because both $A_0$ and $C_0$ are under-estimated in oxRNA. 
Properties which depend on just one of these constants will likely agree less well with experiment.

For low stretching forces, we furthermore observe a torque ``overshoot'' (the increase of torque before reaching the saturated regime with increased superhelical density) upon buckling, as was found experimentally for both DNA~\cite{Forth2008} and RNA.\cite{Lipfert2014} 
This overshoot is due to the need to nucleate the end loop of the plectoneme and its magnitude is set by the difference between the free-energy cost of forming the plectoneme end-loop and the free-energy cost of adding one superhelical turn to an existing plectoneme.\cite{Brutzer2010}
Decreasing the solvent ionic strength and hence increasing the electrostatic strand repulsion is expected to change the free-energy of the relatively large end-loop less than that of additional, more tightly wound plectoneme turns.
Therefore, a reduction of the overshoot with decreasing salt concentration is expected.\cite{Brutzer2010}
Consistently, we observe a smaller overshoot compared to analogous simulations of DNA at 500\,mM monovalent salt concentration.\cite{Matek2015}

The mechanical parameters of our model obtained so far can be used to derive the torsional stiffness of the plectonemic state $P$ by fitting to an analytical model introduced by Marko,\cite{Marko2007} as explained in detail in the Supplementary Material~\cite{Note1} (see Fig.~\ref{Fig:Mechanics}(c)).
While trend and order of magnitude agree, the simulation results deviate from the theoretical prediction due to finite size effects. 
Experimental measurements of Ref.~\onlinecite{Lipfert2014} from a 4.2-kbp dsRNA system show a qualitatively similar deviation from the analytical model, suggesting that at higher forces the postbuckling slopes are slightly higher than predicted by the analytical model.

We summarize the mechanical parameters of dsRNA inferred in this study for the coarse-grained model at a monovalent salt concentration of 100\,mM in Table~\ref{tab:parameters}, along with the corresponding values determined from experiments.
In order to be consistent with common experimental protocols,\cite{Mosconi2009} the equilibrium twist angle $\theta_0$ and the corresponding pitch $p_0=2\pi/\theta_0$ were obtained by demanding that the overall torque $\Gamma (F,\theta)$ exerted on the strand by the traps vanish in a system with that twist angle: $\Gamma(F,\theta_0)=0$.

\squeezetable
\begin{table}
\caption{Summary of mechanical parameters determined for dsRNA at 100-150\,mM monovalent salt concentration.}
\label{tab:parameters}
\begin{tabular*}{\columnwidth}{l l l}
 \hline
  Parameter & oxRNA & experiment\\
  \hline \hline
  Bending persistence length $A_0$ [nm] & 32 & 57-63~\cite{Abels2005,Herrero-Galan2013,Lipfert2014}\\
  Torsional persistence length $C_0$ [nm] & 79 & 100~\cite{Lipfert2014}\\
  Torsional stiffness of plectonemes $P$ [nm]& 22 & 20~\cite{Lipfert2014}\\
  Extension modulus $K$ [pN]& 116 & 350-500~\cite{Lipfert2014,Herrero-Galan2013}\\
  Equilibrium helical pitch $p_0$ [bp] & 11.14 & 10.7-11~\cite{neidle2010principles}\\
  Equilibrium twist angle $\theta_0$ [deg] & 33.3 & 32.7-33.5~\cite{neidle2010principles}\\
  Twist-stretch coupling \\$d\Delta L/d Lk$ [nm/turn]  &-0.72 & -0.85~\cite{Lipfert2014}\\
  \hline
\end{tabular*}
\end{table}

\section{Summary and Conclusions}
We have investigated the mechanical response of dsRNA to twist and stretching force in a coarse-grained computational model. 
To our knowledge, this is the first full determination of the buckling behaviour of dsRNA in a model at single-nucleotide resolution that consistently incorporates the salt-dependent thermodynamics of double strand denaturation.
Reproducing the persistence lengths in a quantitatively accurate fashion has proven more challenging for dsRNA than for dsDNA in the framework of coarse-grained simulations, both for oxRNA,\cite{Sulc2014} as well as in base-pair level models such as the recent work by Chou {\it et al.}.\cite{Chou2014}
This is presumably due to the more complicated structure of the A-form helix in dsRNA as opposed to the B-helix in dsDNA.
However, by comparing to experimental data, we have shown that a physical description of the properties of dsRNA under torsion and tension is still possible.
The experimentally observed decrease in end-to-end distance with increased twist (i.e. positive twist-stretch coupling) of RNA is captured well by oxRNA. 
By contrast, our coarse-grained model of DNA does not reproduce the anomalous (negative) twist-stretch coupling observed in dsDNA.\cite{Ouldridge2011,Matek2014a}
We note that the model of Ref.~\onlinecite{Chou2014} has reported negative twist-stretch coupling for both dsDNA and dsRNA.
This suggests that, although both positive and negative twist-stretch coupling can be represented in the framework of coarse-grained models, capturing the differential behaviour in both molecules may be beyond the scope of present coarse-grained descriptions. \\

Our model is unable to capture the disappearance of the positively supercoiled plectonemic state at higher stretching forces. 
Given the simplified nature of the oxRNA model, it is perhaps not too surprising that we are unable to capture this ``P-RNA'' state, 
however we note that the structure and physical origins of this state are not yet fully understood. 

Similar to our simulations of DNA, we observe plectonemes with denaturation bubbles at the tips of their end-loops for negative supercoiling and intermediate stretching forces of approximately 2\,pN.
This coupling of denaturation and writhing  occurs because the highly bent tip of a plectoneme is a particularly favourable location for the nucleation of a bubble; similarly a bubble is a favourable site at which to initiate writhing.
In contrast to dsDNA, no end-loop denaturations occurred for positive supercoiling up to stretching forces of 6\,pN, presumably due to the stronger binding between Watson-Crick base pairs in dsRNA.
When a plectoneme with a tip bubble is present, we predict it to be preferentially localised in weak parts of the strand sequence, by a mechanism analogous to the one described for dsDNA.\cite{Matek2015}

Summing up, we have presented a comprehensive study of dsRNA under torsional and extensional stress. 
While reproducing the detailed behaviour of the molecule remains a challenge for coarse-grained modelling, our findings are in good agreement with experimental results and provide the basis for capturing the behaviour of more complex RNA structures.

\section*{Acknowledgements}
The authors wish to thank the EPSRC for financial support and Advanced Research Computing, Oxford for computing time. 
We thank Lorenzo Rovigatti and Flavio Romano for their contributions to the development of the oxDNA code, and Jan Lipfert for sharing his data and for useful discussions. 
The donation of GPU cards by the NVIDIA corporation is gratefully acknowledged.\newpage

\bigskip

\newpage
\setcounter{figure}{0}
\setcounter{table}{0} 
\setcounter{section}{0} 
\renewcommand{\thefigure}{S\arabic{figure}}
\renewcommand{\thetable}{S\arabic{table}}
\renewcommand{\thesection}{\Roman{section}}
\onecolumngrid
\newcommand{\celsius}{\,^{\circ}{\rm C}}
\newcommand{\angstrom}{\textup{\AA}}
\newcommand{\sodium}{[Na$^+$]}
\newcommand{\magnesium}{[Mg$^{2+}$]}
\newcommand{\potassium}{[K$^{2+}$]}
\newcommand*{\plimsoll}{{\ensuremath{-\kern-4pt{\ominus}\kern-4pt-}}}

\newpage
\begin{center}
 {\Large \bf Supplementary Material}
\end{center}
\setcounter{figure}{0}
 \makeatletter 
 \renewcommand{\thefigure}{S\@arabic\c@figure}
 \setcounter{equation}{0}
 \renewcommand{\theequation}{S\@arabic\c@equation}
 \setcounter{table}{0}
 \renewcommand{\thetable}{S-\@Roman\c@table}
  \setcounter{section}{0}
 \renewcommand{\thesection}{S-\@Roman\c@section}
\maketitle

\section{Extension of the oxRNA model to include salt dependence}
Following the incorporation of salt-dependent interactions in the oxDNA model of DNA \cite{Snodin2014}, we present here a similar extension of the oxRNA model of Ref.~\onlinecite{Sulc2014} to include salt dependence. 
We parameterise the new interaction in the oxRNA model to reproduce the melting temperatures of RNA duplexes at different monovalent (Na$^+$) salt concentrations. The details of the fitting procedure used can be found in Ref.~\onlinecite{Snodin2014}.

The additional term introduced into the oxRNA potential to capture salt effects is of a modified Debye-H\"uckel form
\begin{equation}
V_{\rm electrostatic}\left(r^{\rm b-b},T,I\right) = \begin{cases}
	V_{\rm DH}(r^{\rm b-b}, T,I)  & \text{if $ r_{\rm smooth} > r^{\rm b-b} $},\\
	V_{\rm smooth} (r^{\rm b-b},T,I) & \text{if $r_{\rm cut} > r^{\rm b-b} \geq r_{\rm smooth} $},\\
	0 & \text{otherwise}.
	\end{cases} 
\end{equation}
where
\begin{equation}
 V_{\rm DH} \left(r^{\rm b-b}, T,I \right) = \frac{ \left( q_{\rm eff} e \right)^2}{4 \pi \epsilon_0 \epsilon_{\rm r}} \frac{\exp\left( - r^{\rm b-b} / \lambda_{\rm DH} \left(T,I\right)  \right)  }{r^{\rm b-b}}
\end{equation}
and 
\begin{equation}
 \lambda_{\rm DH}(T,I) = \sqrt{\frac{\epsilon_0 \epsilon_r k_{\rm B} T}{2N_{\rm A} e^2 I}}.
\end{equation}
$V_{\rm smooth}$ is given by
\begin{equation}
 V_{\rm smooth} = b \left( r^{\rm b-b} - r_{\rm cut} \right)^2 
\end{equation}
with $b$ and $ r_{\rm cut}$ chosen so that $V_{\rm electrostatic}$ is smooth and differentiable.
This truncation of $V_{\rm DH}$ at finite distance $r_{\rm cut}$ allows for much faster calculation of forces and pairwise energies between particles.
We set $r_{smooth}=3\lambda_{\rm DH}$, the same as for the oxDNA2 model,\cite{Snodin2014} where only negligible differences in oligomer melting temperatures were found when using even larger $r_{\rm smooth}$.
In the equations above, $I$ is the molar salt concentration, $e$ is the electron charge, $k_{\rm B}$ is the Boltzmann constant, $N_{\rm A}$ is Avogadro's number, 
$T$ is the temperature, $\epsilon_0$ is the vacuum permittivity and $\epsilon_r$ is the relative permittivity of water (which we set to 80). The distance between the interacting sites, which are placed on the backbone sites of the rigid bodies representing the nucleotides in oxRNA, is denoted as $r^{\rm b-b}$. 

In Debye-H\"uckel theory, $q_{\rm eff}$ is 1. Here, we used the fitting procedure of Ref.~\onlinecite{Snodin2014} to find the optimal value of $q_{\rm eff}$ for the coarse-grained model by fitting it to the melting temperatures of 5, 6, 7, 8, 10 and 12 mers at salt concentrations ranging from $0.1\,{\rm M}$ to $0.5\,{\rm M}$.
The fitting was performed using the average-base oxRNA model, to which $V_{\rm electrostatic}$ had been added. To obtain the melting temperatures of the RNA duplexes to which we fitted the model, 
we use the melting temperatures as predicted by the nearest-neighbour model by Turner {\it et al.}\cite{mathews1999expanded}, where the respective free-energy contribution of each base pair to the duplex stability have been averaged over all possible combinations of Watson-Crick base-pair steps\cite{Sulc2014}. The nearest-neighbour model was derived for $1\,{\rm M}$ salt. To obtain the melting temperatures for lower salt concentration, we correct the free-energy stability of a duplex by adding an extra destabilizing term to the duplex entropy taken from Ref.~\cite{stephenson2013combining}
\begin{equation}
\label{eq_s}
 \Delta S(N,I) = 0.349 N \log\left( I \right) \, \, {\rm cal}\, {\rm mol}^{-1}\, {\rm K}^{-1}
\end{equation}
where $N$ is the number of phosphates and $I$ is the molar salt concentration. A duplex can have phosphates present at both 3' and 5' ends of each strand, but can also have the phosphates cut at one of the ends of each strand.
As our coarse-grained model does not include an explicit representation of the phosphate group, we chose the magnitude of the charges placed on the nucleotides at both the $3^{\prime}$ and $5^{\prime}$ ends of the strand to be $ q_{\rm eff}/2$. 
This choice leads to a total charge on the RNA duplex that will be the same as if the phosphate charges were cut at one of the ends.
Thus, it should be kept in mind that the oxRNA model cannot reproduce subtleties caused by having the phosphates cut off one or both ends.

The correction to the entropy contribution for the nearest-neighbour model in Eq.~\ref{eq_s} is based on the hairping unzipping experiments in Ref.~\onlinecite{stephenson2013combining}, where the stability of a hairpin was obtained for varying salt concentrations and temperatures. The average destabilization free-energy was observed to be 
$-0.054 N \log\left( I \right) {\rm kcal}/{\rm mol}$, which is similar to that for DNA duplexes at $37\celsius$,
which is $\Delta G_{37} = -0.057 N \log\left( I \right)$ according to Ref.~\onlinecite{SantaLucia2004}. 
In the nearest-neighbour model for DNA melting in Ref.~\onlinecite{SantaLucia2004}, this destabilisation is taken to be only of entropic origin. We hence interpreted the destabilization derived from the RNA hairpin unzipping experiments also as contributing to the entropy in the nearest-neighbor model for RNA thermodynamics. If, however, a more detailed study of RNA duplex or hairpin thermodynamics at varying salt concentrations becomes available, we might need to revisit our parametrization and fit it to more accurate estimations of melting temperatures at varying salts.

We obtained $q_{\rm eff}$ equal to $1.26$ from the fitting procedure. 
We note the resulting $q_{\rm eff}$ is larger than 1, but given the complexity of potential salt effects, and the simplicity of our mean-field Debye-H\"uckel representation, not too much can be read into these numerical values.

To test the fitted value of $q_{\rm eff}$, we studied the melting temperatures of several RNA duplexes at varying salt concentrations with virtual-move Monte Carlo simulations (VMMC), using the variant from the Appendix of Ref.~\onlinecite{Whitelam2009}.  Each simulation was run for at least $3\times10^{11}$ steps. The results are shown in Table \ref{table_averagedmodel} for the average-base oxRNA model with the new salt dependence included.

\begin{table}
 \centering
 \begin{tabular}{c | c | c | c}
   & \multicolumn{3}{c}{\bf{Salt concentration [M]}} \\
  \hline {\bf Motif}  &  $0.1$ &  $0.3$ & $0.5$  \\ \hline 
  6-mer &  $26.3\, (1.3)$ / $25.9$  & $31.2 \,(2.0)$ / $29.7$ & $33.6 \, (0.9)$ / $31.4$   \\ 
  8-mer & $44.9 (4.0)$ / $46.6$  & $52.9 (1.3)$ / $50.7$ & $54.2 (0.4)$ / $52.6$  \\
  10-mer &  $54.6 (2.0)$ / $58.5$ & $65.4 (2.1)$ / $62.8$ & $65.4 (0.5)$ / $64.8$  \\
 \hline
 \hline
 \end{tabular}
\caption{ The melting temperatures of RNA duplexes at different salt concentrations for the average-base parametrization of oxRNA with the new salt-dependent term included ($T_m$) compared to the melting temperatures of the averaged nearest-neighbour model with the salt correction as introduced in Eq.~\ref{eq_s} ($T_m({\rm NN^{avg}})$). 
The individual cells in the table are in the form $T_m$ (error) / $T_m({\rm NN^{avg}})$, where the error was calculated as the standard deviation of the melting temperatures estimated from 5 different independent simulations. 
The melting temperatures $T_m$ were estimated from VMMC simulations and are for a strand concentration of $4.2 \times 10^{-5}\,\rm{ M}$. \label{table_averagedmodel}
}
\end{table}

\section{Boundary conditions}
\begin{figure}[h!]
 \includegraphics[width=.9\columnwidth]{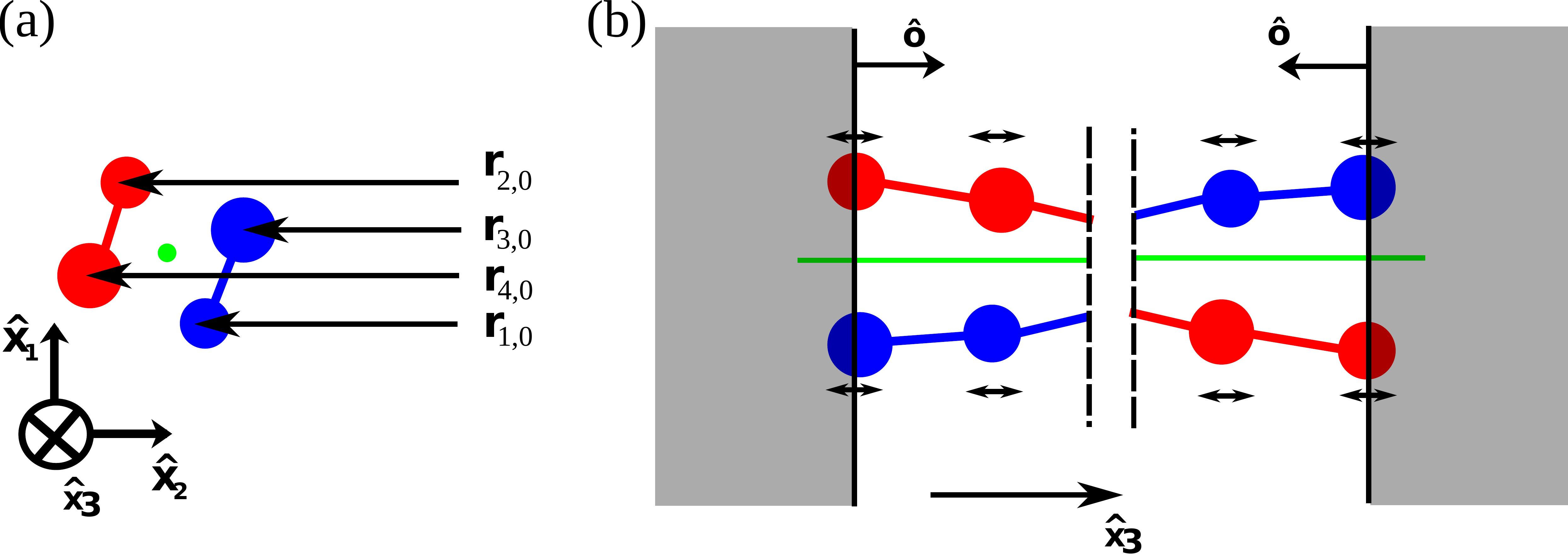}
 \caption{Schematic depiction of the boundary conditions used, illustrated for the last 2\,bp at each end of the dsRNA system: (a) View along the double strand axis. 5 nucleotides on each strand end are constrained by 2-dimensional harmonic traps, which fix the boundary nucleotides to positions $\textbf r_{n,0}$ in planes perpendicular to the strand axis (green). (b) View perpendicular to the double strand. Due to the 2-dimensional traps, nucleotides are unconstrained only in the strand-axis direction. A repulsion plane perpendicular to the strand axis is tagged to the last base pair. Movement of nucleotides into the area below the end base pair (shaded grey) is therefore excluded. In order to allow unconstrained strand extensibility, the repulsion plane does not act on the first two base pairs along the strand.}
 \label{fig:boundary_schematic}
\end{figure}

In order to keep the superhelical density in a dsRNA double strand constant during a simulation, 5 base pairs were added to the 600\,bp-system at each end, and constrained in stiff, two-dimensional harmonic traps.
These traps only exert forces in the plane perpendicular to the setup axis of the double strand, thus not causing any linear elongation of the system.
Analogous constraining boundary conditions have been successfully used before in simulations of cruciform extrusion~\cite{Matek2012} and dsDNA plectoneme structures~\cite{Matek2015}.
A schematic overview of the boundary conditions applied is shown in Fig.~\ref{fig:boundary_schematic}.

The two-dimensional harmonic traps used to keep the superhelical density of the system constant are implemented by a potential of the form
\begin{equation}
 V_{\rm trap}({\textbf r_{n}} ; {\textbf r_{n,0}}) = \frac{1}{2} \sum_{i=1}^3 k_{\rm trap}^i (r^i_{n}-r_{n,0}^i)^2,
\end{equation}
where ${\textbf r_n = (r^1_n, r^2_n, r^3_n)}$ is the centre-of-mass position of the $n$-th trapped nucleotide and the corresponding trap position is ${ \textbf r_{n,0}=(r^1_{n,0}, r^2_{n,0}, r^3_{n,0})}$, chosen initially such as to fix a given twist angle of the strand.
We found that choosing $k_1^{\rm trap}=k_2^{\rm trap}=58.7$N/m and $k_3^{\rm trap}=0$ kept the superhelical density fixed by preventing rotations of the 5-bp handles at the double strand ends, while not hindering strand extension along the setup axis $\mathbf{\hat{x}_3}$.

The RNA duplexes studied in this work have finite length, which means that more distant parts of the system can pass around the strand ends. 
Such a process would modify the superhelical density $\sigma$ of the system.
Therefore, such movements of the system are prevented in our simulations by repulsion planes oriented perpendicular to the setup axis $\mathbf{\hat{x}_3}$ which co-move with the first boundary nucleotide of the two single RNA strands in the system.
Repulsion planes generate a potential 
\begin{equation}
 V_{\rm plane}({\textbf r};{\textbf R}) = \frac{1}{2} k^{\rm plane} \left( \left( {\textbf r}-{\textbf R} \right) \cdot \mathbf{\hat{o}} \right)^2 \theta(-\left( {\textbf r}-{\textbf R} \right) \cdot \mathbf{\hat{o}}),
\end{equation}
where {\bf r} is the centre-of-mass position of an affected particle, {\bf R} and $\mathbf{\hat{o}}$ are anchor point and orientation of the plane, and $\theta$ is the Heaviside step function.
We choose $\mathbf{\hat{o}}=\mathbf{\hat{x}_3}$ and $\mathbf{\hat{o}}=-\mathbf{\hat{x}_3}$ for the lower and upper repulsion planes respectively, and set ${\textbf R}$ equal to the instantaneous positions of the first and last double strand boundary base pair.
To avoid restricting free strand extensibility in the $\mathbf{\hat{x}_3}$ direction, the repulsion planes are set up to not interact with the next-to-last boundary base pairs at both strand ends.
In all simulations, we chose parameters $k^{\rm plane}=29.3$\,pN/nm, which prevented the duplex from passing around its ends during all simulation runs.

\section{Determining extensional properties of dsRNA}
\begin{figure}[h!]
 \includegraphics[width=.9\columnwidth]{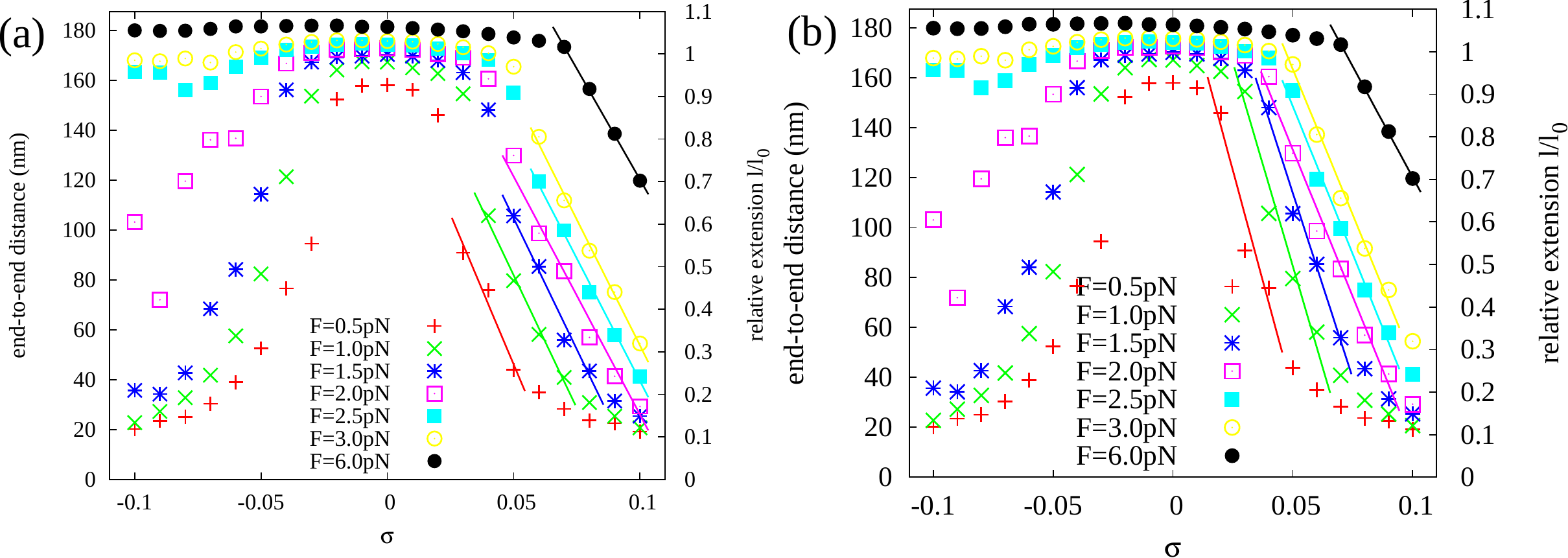}
 \caption{Mean strand end-to-end extensions for all values of parameters $F$ and $\sigma$ studied in this work. Postbuckling slopes were determined by fitting linear functions to the overtwisted branches of the hat curves. As the slopes obtained somewhat depend on the fit range chosen, we performed two different fits over slightly different ranges of $\sigma$, as shown in (a) and (b). Using a thermodynamic model due to Marko~\cite{Marko2007}, the stiffness of the plectonemic phase can be determined from this post-buckling slope.}
 \label{fig:full_hatcurve_fitting}
\end{figure}

The full ``hat curves'' for all stretching forces $F = 0.5, 1.0, 1.5, 2.0, 2.5, 3.0 $ and $6.0$\,pN and superhelical densities $-0.10 \leq \sigma \leq +0.10$ are shown in Fig.~\ref{fig:full_hatcurve_fitting}.
In order to measure the decrease of end-to-end extension as a function of added superhelical density in the post-buckling regime, we fitted linear functions to the overtwisted branch of the hatcurves (see Fig.~\ref{fig:full_hatcurve_fitting}).
For low stretching forces, the postbuckling slope of the hat curves decreases at high levels of supercoiling. 
As has been noted before~\cite{Salerno2012a,Brutzer2010,Matek2015}, this finite-size effect is due to the interactions of the double strand with the system boundaries.
In order to obtain the generic behaviour of the system, we attempted to restrict the fitting to a range in $\sigma$ in which the postbuckling curve exhibits no non-linearities (see Fig.~\ref{fig:full_hatcurve_fitting}).
There is some ambiguity in choosing the range of the linear fits. We therefore performed two separate fits where we shifted the fitting domain by one point towards the buckling transition, as shown in Fig.~\ref{fig:full_hatcurve_fitting}.
The values shown in Fig.~3(c) of the main paper refer to the mean and standard deviation of the two values obtained in this way.

The slopes thus determined can then be directly compared to experimental results, as shown in Fig.~3(c) of the main text. 

The measured values of the post-buckling slopes can furthermore be used to determine the twist stiffness of the plectonemic state $P$ by fitting to a relation obtained by Marko~\cite{Marko2007}.
Following Refs.~\onlinecite{Lipfert2014} and \onlinecite{Forth2008}, this relation is:
\begin{equation}
 \frac{d \Delta L}{d \Delta Lk} = \frac{p_0\left[ 1- \frac{1}{2}\sqrt{\frac{k_BT}{A_0F}}-\frac{\theta_0^2 C_0^2}{16}\left( \frac{k_BT}{A_0F}\right)^{3/2} \left( \frac{1}{c} \sqrt{\frac{2pg}{1-p/c}}\right)^2 \right]}{\sqrt{\frac{2pg}{1-p/c}}\left(\frac{1}{p}-\frac{1}{c} \right)},
 \label{eqn:Marko_plectoneme_stiffness}
\end{equation}
where $A_0$ and $C_0$ are the bending and twist persistence lengths, $p_0$ is the equilibrium helical pitch, $\theta_0$ the equilibrium twist angle and $F$ the stretching force.
Furthermore, $g = F - \sqrt{Fk_BT/A_0}$, while $p=k_BT P \theta_0^2$ and $c = k_BT C \theta_0^2$ are proportional to $P$ and $C_0$ respectively.
The result of fitting Eq.~\ref{eqn:Marko_plectoneme_stiffness} to the postbuckling slopes determined from simulations is shown in Fig.~3(c) of the main text. 

\section{Detection of double strand melting and plectoneme position}
As the value of $V_{\rm HB}$ paired nucleotides assumes continuous values, it is necessary to define a cutoff criterion to determine whether a given pair of nucleotides is base-paired or not.
Following the approach taken previously~\cite{Sulc2014,Matek2015}, we counted a base-pair as formed when the interaction energy from hydrogen bonding between two nucleotides was below $-4.13 \times 10^{-21}$\,J, corresponding to approximately $15\%$ of the typical energy of a fully formed hydrogen bond.

In order to assign a position variable to a given plectoneme structure, we used the plectoneme detection algorithm described in detail in Ref.~\onlinecite{Matek2015}.
The individual steps of plectoneme detection are~\cite{Matek2015}:
\begin{itemize}
 \item Start from a double strand end, loop over all base pair centre points
 \begin{itemize}
  \item If any part of the remaining double strand that is more than $N_c$ bp away along the contour of the duplex has a distance $d_{\rm lin}<d_{\rm lin}^{0}$, record the index of the current base pair as the beginning of a plectoneme, if the beginning of a plectoneme has not yet been detected before.
  \item If $d_{\rm lin}>d_{\rm lin}^{0}$ for all base pair centres of the remaining double strand and a plectoneme beginning has been detected before, record the current base pair index as the end of a plectonemic region and continue searching for further plectonemes from the next base pair centre
 \end{itemize}
 \item The plectoneme position is the mean between the base pair indices of the beginning and end of a plectonemic region
 \item The plectoneme size is the difference between the base pair indices of the beginning and end of a plectonemic region
\end{itemize}

The systems studied in the present work are simulated at a monovalent ionic strength of 100\,mM, which is significantly lower than the 500\,mM ionic strength considered for the analogous dsDNA system in Ref.~\onlinecite{Matek2015}.
As a consequence of the increased electrostatic strand repulsion due to lower salt, the diameter of the end-loop and plectoneme stem are expected to slightly increase.
It was found that the properties of these somewhat larger structures is best captured when setting the detector parameters to $d_{\rm lin}^{0}=10.1$\,nm and $N_c=50$\,bp, which are slightly larger than the values used in Ref.~\onlinecite{Matek2015}.
We note that $N_c$ represents a lower limit on the size of plectoneme structures that can be detected using the detection algorithm outlined above.
However, at an ionic strength of 100\,mM, typical plectoneme structures are significantly larger than 50\,bp, and are therefore reliably detected by the algorithm.

As in our previous work on dsDNA (Ref.~\onlinecite{Matek2015}), a tip-bubble plectoneme is defined as a plectoneme whose midpoint as defined by the detection algorithm is less than 20\,bp away from the centre of a denaturation bubble.

\section*{Bibliography}

\bibliographystyle{aipnum4-1}
\bibliography{RNA_references}
\end{document}